\documentclass[reqno]{amsart}
\usepackage{amsmath,amsfonts,amssymb}
\usepackage{verbatim}
\usepackage{enumerate}
\usepackage{url}
\def\N{\mathbb N}

\def\Z{\mathbb Z}

\def\F{\mathbb F}
\def\ord{\mathop{\rm ord}\nolimits}

\def\lcm{\mathop{\rm lcm}}

\theoremstyle{plain}
\newtheorem{theorem}{Theorem}[section]
\newtheorem{lemma}[theorem]{Lemma}
\newtheorem{definition}[theorem]{Definition}
\newtheorem{corollary}[theorem]{Corollary}

\newtheorem{remark}[theorem]{Remark}

\def\proof{{\it Proof: }}
\def\qed{\hfill\hbox{$\square$}}

\theoremstyle{definition}

\numberwithin{equation}{section}

\author[F.E. Brochero Mart\'{\i}nez]{F. E. Brochero Mart\'{\i}nez}
\address{
Departamento de Matem\'{a}tica\\
Universidade Federal de Minas Gerais\\
UFMG\\
Belo Horizonte, MG\\
 30123-970\\
 Brazil\\
 }
 \email{fbrocher@mat.ufmg.br }

\date{\today
}

\title[Number of minimal cyclic codes]{Number  of minimal  cyclic codes  with given   length and dimension}

\begin{document}
\maketitle

\begin{abstract} 
In this article, we count the quantity of  minimal cyclic codes of length $n$ and dimension $k$  over  a finite field $\F_q$,   in the case  when the prime factors of $n$ satisfy a special condition. 
This problem is equivalent to count the quantity of irreducible factors  of $x^n-1\in \F_q[x]$ of degree $k$.
\end{abstract}

\section{Introduction}
 Let $\F_q$ be a finite field with $q$ elements.  A linear $[n,k;q]$ code $\mathcal C$   is a linear subspace of $\F_q^n$ of dimension $k$.
 $\mathcal C$ is called a {\em cyclic code} if $\mathcal C$ is invariant by a shift permutation, i.e., if $(a_0,a_1,\dots, a_{n-1})\in \mathcal C$ then $(a_{n-1},a_0,a_1,\dots, a_{n-2})\in \mathcal C$. 
  It is known that every cyclic code can be seen as an ideal of the ring $\frac {\F_q[x]}{(x^n-1)}$. In addition, since $\frac {\F_q[x]}{(x^n-1)}$ is a principal ring,  every ideal is generated by a polynomial $g(x)$ such that $g$ is a divisor of $x^n-1$. Thus, the polynomial $g$ is called {\em generator} of the code and the polynomial $h(x)=\frac {x^n-1}{g(x)}$ is called the {\em parity-check} polynomial of $\mathcal C$.  
 Observe that $\{g, xg,\dots x^{k-1}g\}$, where $k=\deg(h)$, is a basis of the linear space $(g)\in \frac {\F_q[x]}{(x^n-1)}$, then  the dimension of the code is the degree of the parity-check polynomial. 
A cyclic code $C$ is called  {\em minimal cyclic code} if $h$ is an irreducible polynomial in $\F_q[x]$. 
Thus, the number of irreducible factors of $x^n-1\in \F_q[x]$  corresponds  to the number of minimal cyclic codes of length $n$ in $\F_q$.  
Specifically, there exists a bijection between  the minimal cyclic codes of dimension $k$ and length $n$ over $\F_q$, that we denote by $[n,k;q]$,  and the irreducible factors of $x^n-1\in \F_q[x]$ of degree $k$.

Irreducible cyclic codes are very interesting by its applications in communication, storage  systems  like compact disc players, DVDs, disk drives,  two-dimensional bar codes, etc. (see \cite[Section 5.8 and 5.9]{FaCa}).
  The advantage of the cyclic codes, with respect to other linear codes,  is that they have efficient encoding and decoding algorithms (see  \cite[Section 3.7]{FaCa}). For these facts, cyclic codes have been studied for the last decades  and many progress has been found (see \cite{HuPl}).

A natural question is how many minimal  cyclic codes of length $n$ and dimension $k$ over $\F_q$ does there exist? In other words, the quations is: given $n$, $k$ and $\F_q$, find an explicit formula for the number of minimal cyclic $[n,k;q]$-codes. 
  This question is in general unknown, and   how  to construct all of them too. 

In this article, we determine the number of minimal cyclic $[n,k;q]$-codes assuming  that the order of $q$ modulo  each prime factor of $n$   satisfies   some special relation. 

\section{Preliminaries }
Throughout this article, $\F_q$ denotes a finite field of order $q$, where $q$ is a power of a 
 prime.  For each $a\in \F_{l}^*$,  $\ord(a)$ denotes the order of $a$ in a multiplicative group $\F_{l}^*$, i.e. $\ord(a)$ is the least positive integer $k$ such that $a^k=1$.  In the same way, we denote by $\ord_n b$, the order of $b$ in a multiplicative group $\Z_n^*$ and 
 $\nu_p(m)$ is the maximal power of $p$ that divides $m$.
In addition, for each irreducible polynomial $P(x)\in \F_q[x]$,  $\ord(P(x))$ denotes the order of some root of $P(x)$ in some extension of $\F_q$.   

It is a classical result (see, for instance, \cite{But})  to  determine the number of factors of $x^n-1$ and its degree, when the order is given. 


\begin{theorem} Let $n$ be a positive integer such that $\gcd(n,q)=1$, then each factor of $x^n-1\in \F_q[x]$  has order $m$, where $m$ is a divisor of $n$. In addition, for each $m|n$, there exist $ \frac {\varphi(m)}{\ord_m q}$ irreducible factors and each of these factors has degree $\ord_m q$. 
\end{theorem}

As  a consequence of this  theorem (see proposition  2.1 in \cite{Ago}),
the number of factors of degree $k$ of $x^n-1$ is $\sum\limits_{ m|n \atop \ord_m q=k}\frac{ \varphi(m)}k$ and then the total number of irreducible factors is  $\sum\limits_{ m|n} \frac{ \varphi(m)}{\ord_m q}$. 
So, the number of irreducible factors of degree $k$  is zero if any $m$ divisor of $n$ satisfies $\ord_m q=k$.  
Clearly, this formula is not really explicit, because it depends on the calculation of the orders $\ord_m q$ for every divisor of $n$. 

 An equivalent approach  is to use  the technique  of  $q$-cyclotomic classes (see \cite{Xam} page 157 or \cite{MaSl} Chapter 8). In fact,  the $q$-cyclotomic  class of $j$ modulo $n$ is   the set
$\{j, jq, jq^2,\dots, jq^{k-1}\}$
whose elements are distinct  modulo $n$ and $jq^k\equiv  j\pmod n$.   This $q$-cyclotomic class determines one irreducible factor of $x^n-1$ of degree $k$.

If we denote by
 $\mathcal C_k $    the set of numbers $j$, with  $1\le j\le n$ that have $q$-cyclotomic class with $k$ elements, then
\begin{align*}
\mathcal C_k&=\{j\le n; \text{ $k$ is the minimum positive integer  such that $jq^k\equiv j\pmod n$}\}\\
&=\left\{j\le n; \text{ $k$ is the minimum positive integer  such that $q^k\equiv 1\pmod {\frac n{\gcd(n,j)}}$}\right\}\\
&=\left\{j\le n; \ k=\ord_{\frac{n}{\gcd(n,j)}} q\right\}.
\end{align*}
Since each $q$-cyclotomic class determines a minimal cyclic code, then 
the number of minimal cyclic $[n,k;q]$-codes is $\dfrac {|\mathcal C_k|}k$.

Using this technique, in \cite{SaSe} and  \cite{KuAr}  are  shown   explicit formulas for the total of minimal cyclic codes for some special cases.

\begin{theorem}[\cite{SaSe}]\label{teo2.2}
Suppose that $n=p_1^{\alpha_1}p_2$  satisfies that 
$d = \gcd(\varphi (p_1^{\alpha_1} ), \varphi(p_2))$, $p_1\nmid (p_2 - 1)$ and $q$ is a primitive root $\mod p_1^{\alpha_1}$ as well as $\mod p_2$. Then the number of minimal cyclic codes of length $n$ over $\F_q$ is $\alpha_1(d + 1) + 2$.
\end{theorem}

\begin{theorem}[{\cite[Theorem 2.6]{KuAr}}]\label{teo2.3} Suppose that  $n=p_1^{\alpha_1}\cdots p_l^{\alpha_l}$ satisfies that  $\ord_{p_j^{\alpha_j}}q=\varphi(p_j^{\alpha_j})$  for every $j$, and 
$\gcd(p_j-1, p_i-1)=2$ for every $i\ne j$. Then the number of minimal cyclic codes of length $n$ over $\F_q$ is
$$\frac {(2\alpha_1+1)(2\alpha_2+1)\cdots (2\alpha_k+1)+1}2.$$
\end{theorem}

Besides, some explicit formulas for the number of $[n,k;q]$-codes for some particular values of  $n$ and $q$ are known

\begin{theorem}[{\cite[Corollary 3.3 and 3.6]{BGO} }]\label{teo2.4}
Suppose that $n$ and $q$ are numbers such that every prime factor of $n$ divides $q-1$.  Then
\begin{enumerate}
\item If $8\nmid n$ or $q\not \equiv 3 \pmod 4$ then the number of minimal cyclic $[n,d;q]$-codes  is
$$\begin{cases}
\frac{\varphi(d)}d\cdot \gcd(n,q-1)&\text{if $d\mid \frac {n}{\gcd(n,q-1)}$}\\
 0 &\text{otherwise}
\end{cases}
$$
The total number of minimal cyclic codes of length $n$ is 
$$ \gcd(n,q-1)\cdot\prod_{p|m\atop p{\text{ prime}}} \left(1+\nu_p(m)\frac {p-1}p\right),$$
where $\varphi$ is the Euler Totient function.
\item If $8| n$ and $q \equiv 3 \pmod 4$ then the number of minimal cyclic $[n,d;q]$-codes  is
$$
\begin{cases}
\frac{\varphi(d)}d\cdot \gcd(n,q-1)&\text{if $d$ is odd and $d\mid \frac {n}{\gcd(n,q^2-1)}$}\\
\frac{\varphi(k)}{2k}\cdot (2^r-1)\gcd(n,q-1)&\text{if $d=2k$, $k$ is odd and $k\mid \frac {n}{\gcd(n,q^2-1)}$}\\
\frac{\varphi(k)}{k}\cdot 2^{r-1} \gcd(n,q-1)&\text{if $d=2k$, $k$ is even and $k\mid \frac {n}{\gcd(n,q^2-1)}$}\\
 0 &\text{otherwise}
\end{cases}
$$
where $r=\min\{\nu_2(n/2),\nu_2(q+1)\}$.  The total number of minimal cyclic codes of length $n$ is 
$$ \gcd(n,q-1)\cdot \left(\frac 12+2^{r-2}(2+\nu_2(m ))\right)\cdot\prod_{p|m\atop p{\text{ odd prime}}} \left(1+\nu_p(m)\frac {p-1}p\right).$$
\end{enumerate}
\end{theorem}

\section{Codes with  power of a prime  length}

In this section, we are going to suppose that $n$ is a power of a prime. In order to determine the number of irreducible codes of length $n$, we need the following lemma, that it is pretty well-known in the Mathematical Olympiads folklore and  it is attributed to
E. Lucas  and R. D. Carmichael (see  \cite{Car}). 

\begin{lemma}[Lifting-the-exponent Lemma]\label{LEL}
Let $p$ be a prime. For  all $a,b\in \Z$  and   $n\in \N$,  such that $p\nmid ab$ and $p|(a-b)$, the following proprieties  are satisfied
\begin{enumerate}[(i)]
\item If $p\ge 3$, then $\nu_p(a^n-b^n)=\nu_p(a-b)+\nu_p(n)$.
\item If $p=2$ and   $n$ is odd  then $\nu_2(a^n-b^n)=\nu_2(a-b)$.
\item If $p=2$ and $n$ is even  then $\nu_2(a^n-b^n)=\nu_2(a^2-b^2)+\nu_2(n)-1$.
\end{enumerate} 
\end{lemma}

As a consequence of the previous  lemma   we obtain

\begin{corollary}\label{order} Let $p$ be a prime and  $\rho=\ord_p q$. 
\begin{enumerate}
\item If $q\not\equiv 3\pmod 4$  or $p\ne 2$   then
$$
\ord_{p^{\theta}} q=\begin{cases} 1 &\text{if $\theta=0$}\\
\rho&\text{if $\theta\le \beta$}\\
\rho p^{\theta-\beta}&\text{if $\theta> \beta$.}
\end{cases}
$$
where  $\beta=\nu_p(q^\rho-1)$.
\item If $q\equiv 3 \pmod 4$  and $p=2$,  then
$$\ord_{2^{\theta}} q=\begin{cases} 1 &\text{if $\theta=0$ or $1$.}\\
2&\text{if $\theta\le \beta$}\\
 2^{\theta-\beta+1}&\text{if $\theta> \beta$.}
\end{cases}
$$
where $\beta=\nu_2(q^2-1)$.
\end{enumerate}
\end{corollary}
\proof
 {\em (1)}  Clearly,  $\ord_{p^\theta} q=\rho$  if $1\le \theta \le \beta$.  In the case $\theta>\beta$, since $\ord_p q$ divides $\ord_{p^\theta}$ then, by Lemma \ref{LEL} item {\em (i)}, we have
$$\theta=\nu_p(q^k-1)=\nu_p(q^\rho-1)+\nu_p\left(\frac k\rho\right)=\beta+\nu_p\left(\frac k\rho\right).$$
In addition to the minimality of  $k$, we obtain that  $\frac k\rho=p^{\theta-\beta}$.

The proof of part {\em (2)} is similar by using  items  {\em (ii)} and {\em (iii)} of Lemma \ref{LEL} .
\qed

\begin{theorem}\label{teo3.2}
Suppose that $n=p^\alpha$, where $p$ is  a prime and $\rho$ and $\beta$ as in the previous lemma. Then
\begin{enumerate}
\item If $p\ne 2$  or $q\not\equiv  3\pmod 4$ then the number of minimal cyclic  $[n,d;q]$-codes is
$$\begin{cases}
\gcd(n, q-1) &\text{if $d
= 1$ }\\
\frac {p^{\min\{\alpha,\beta\}}-1}\rho &\text{if $d=\rho\ne 1$ }\\
\frac{p^{\beta}-p^{\beta-1}}{\rho} & \text{if $d=\rho \cdot p^j$ and $1\le j\le \alpha-\beta$}\\
0&\text{otherwise}
\end{cases}
$$
\item If $n=2^\alpha$   and $q\equiv  3\pmod 4$ then the number of minimal  cyclic $[n,d;q]$-codes is
$$\begin{cases}
2&\text{if $d=1$}\\
1&\text{if $d=2$ and $\alpha=2$}\\
3 &\text{if $d=2$ and $\alpha\ge 3$}\\
2 & \text{if $d=2^j$ and $2\le j\le \alpha-2$}\\
0&\text{otherwise}
\end{cases}
$$
\end{enumerate}
\end{theorem}

\proof   
{\em (1)}   In the case when $k=1$, the number of $[n,1:q]$-codes is equivalent to the number of roots of  the polynomial $x^n-1$ in $\F_q^*$. Since every element of $\F_q^*$ is root of $x^{q-1}-1$,  and $\gcd(x^n-1,x^{q-1}-1)=x^{\gcd(n,q-1)}-1$, we conclude that the number of minimal $[n,1;q]$-codes is $\gcd(n, q-1)$.

Now, suppose that $d\ne 1$. Since $\rho$ divides $\ord_{p^s} q$ for every $s\ge 1$ and $\frac{\ord_{p^s} q}{\rho}$ is a power of  $p$, 
it follows that  if  $\frac k{\rho}$ is not a power of $p$, then there not exist $[n,k;q]$-codes.

In the case when $d=\rho$, by Corollary \ref{order}, we know that $\ord_{p^s} q=\rho$ if and only if $1\le s\le \beta$ and then the number of $[n,\rho;q]$-codes is
$$\sum_{s=1}^{\min\{\alpha,\beta\}} \frac {\varphi (p^s)}{\rho}=\sum_{s=1}^{\min\{\alpha,\beta\}} \frac {p^s-p^{s-1}}{\rho}=\frac {p^{\min\{\alpha,\beta\}} -1}{\rho}$$

Finally, in the case $d=\rho p^j$,  since   $\ord_{p^s}q=\rho p^j$ if and only if $s=j+\beta$, and $s\le \alpha$, we conclude that $j\le \alpha-\beta$ and the number of $[n,\rho p^j; q]$-codes is
$$\frac {\varphi(p^s}{\ord_{p^s}q}=\frac{\varphi(p^{j+\beta})}{\rho p^j}=\frac {p^\beta-p^{\beta-1}}\rho.$$
So, this identity  concludes the proof of (1). 

We note that the proof of (2) is essencially the same of (1) and we omit.
\qed

\begin{remark}  In \cite{BrGi}, we  show one way to  construct the primitive idempotents  of the ring $\frac {\F_q[x]}{(x^n-1)}$ where $n=p^{\alpha}$ and it is known that each  primitive idempotent is a generator of one minimal cyclic code of length $n$.
\end{remark}

\section{The number of cyclic codes given an special condition}
Throughout this section, 
 $n=p_1^{\alpha_1}\cdots p_l^{\alpha_l}$ is  the factorization in primes of $n$, where $n$ is odd or $q\not \equiv 3\pmod 4$. 
Moreover, we put   $\rho_i=\ord_{p_i} q$ and 
 $\beta_i=\nu_{p_i}(q^{\rho_i}-1)$. 


\begin{definition} The pair  $(n,q)$ satisfies the {\em homogeneous order condition} (H.O.C.) if  $\gcd(\rho_i, n)=1$,  for every $i$, and there exists $\rho\in \N$ such that $\rho=\gcd (\rho_i,\rho_j)$,  for every $i\ne j$. 
\end{definition}

Observe that  every pair $(n,q)$ considered in Theorems \ref{teo2.2}, \ref{teo2.3}, \ref{teo2.4} and \ref{teo3.2} satisfies H.O.C.. Furthermore, 
if $(n,q)$ satisfies H.O.C then 
$$R:=\lcm(\rho_1,\rho_2,\dots,\rho_k)=\frac {\rho_1\rho_2\cdots \rho_k} {\rho^{k-1}}$$ and, by  Lemma \ref{LEL}, we have
$$\nu_{p_i}(q^R-1)=\nu_{p_i}( q^{\rho_i}-1)+\sum_{1\le j\le k \atop j\ne i}\nu_{p_i}\left(\frac {\rho_j}{\rho}\right)=\beta_i.$$


\begin{lemma}\label{HOC}
Let $(n,q)$ be a pair which satisfies H.O.C.  and $d=p_1^{\theta_1}\cdots p_l^{\theta_l}$  be a divisor of $n$ other than $1$. Then
$$\ord_{d} q=\frac {\rho d}{\gcd(d, q^R-1)} \prod_{1\le i\le l\atop \theta_i\ne 0} \frac{\rho_i}{\rho}.$$ 
\end{lemma}

\proof 
Observe that if 
 $\theta_i\ne 0$ then
$$\ord_{p_i^{\theta_i}} q=\rho_i \frac {p_i^{\theta_i}}{\gcd(p_i^{\theta_i},q^{\rho_i}-1)}=
\rho_i \frac {p_i^{\theta_i}}{\gcd(p_i^{\theta_i},q^{R}-1)}.
$$
Thus,  in the case when  $d=p_{i_1}^{\theta_{i_1}}\cdots p_{i_s}^{\theta_{i_s}}$, where $\theta_{i_j}\ne 0$, we have
\begin{align*}
\ord_d q&= \lcm (\ord_{p_{i_1}^{\theta_{i_1}}} q, \dots \ord_{p_{i_s}^{\theta_{i_s}}} q)\\
&=\rho \cdot \lcm \left(\frac {\ord_{p_{i_1}^{\theta_{i_1}} }q}\rho, \dots ,\frac {\ord_{p_{i_s}^{\theta_{i_s}}} q}{\rho}\right)\\
&=\rho \prod_{j=1}^s \frac {\rho_{i_j}}{\rho} \frac {p_{i_j}^{\theta_{i_j}}}{\gcd( p_{i_j}^{\theta_{i_j}}, q^R-1)}\\
&=\rho \frac d{\gcd(d, q^R-1)} \prod_{p_i|d} \frac {\rho_i}{\rho}.
\end{align*} \qed

\begin{corollary}\label{dimension} Let $(n,q)$ be a pair which satisfies H.O.C.. 
 If there exist  minimal  cyclic $[n,k;q]$-codes then   
\begin{enumerate}
\item $\gcd(k, \rho_i)=1$ or $\rho_i$, for every $i$.
\item If $p_i$  divides $\gcd(n,k)$, then $\rho_i$ divides $k$.
\item $\gcd(n,k)$ divides $\dfrac {\cdot n}{\gcd(n,q^R-1)}$.
\end{enumerate}
\end{corollary}

\begin{theorem} Let $\F_q$ be  a finite field  and $n$ be a positive integer such that the pair $(n,q)$ satisfies H.O.C. 
and suppose that $n$ is odd or $q\not \equiv 3 \pmod 4$. Let $k$ be a positive integer satisfying the conditions of the corollary \ref{dimension}.
Then the number of  minimal cyclic $[n,k;q]$-codes is
$$
\begin{cases}
\gcd(n,q-1)  & \text{ if $k=1$}\\
\gcd(n, q^R-1)\frac {\varphi(\gcd(k,n))} k
&\text{ if $k\ne 1$}.
\end{cases}
$$
The total number of minimal cyclic codes of length $n$ is 
$$ \frac{\rho-1+\prod\limits_{i=1}^l\left(\frac {\rho}{\rho_i}\left(\varphi(p_i^{\beta_i})
\max\{\alpha_i-\beta_i, 0\} 
 +p^{\min\{ \alpha_i, \beta_i\}} -1\right) +1\right)}{\rho} $$

\end{theorem}

\proof  We are going to suppose that $k\ne 1$,  because the case $k=1$  has been proved in Theorem \ref{teo3.2}. Let $\mathcal I$ be the set of indices $i$ such that $\frac {\rho_i}\rho$ divides $k$, $\mathcal J=\{i\in \mathcal I| p_i$ divides $k\}$ and $\mathcal I_0=\mathcal I\setminus \mathcal J$.

Let $d$ be a divisor of $n$ such that $\ord_d q=k$. By Lemma  \ref{HOC}, it follows that $d|n_{\mathcal I}$ and 
$k=tR_{\mathcal I}$ where 
$$t=\gcd(k,n)=\dfrac d{\gcd(d, q^R-1)}\quad\text{ and }\quad R_{\mathcal I}=\rho\prod_{i\in \mathcal I} \frac {\rho_i}{\rho} .$$ 
Since $t=\prod_{i\in \mathcal I}  p_i^{\theta_i}$, then
$$\theta_i=\nu_{p_i}(d)-\min\{\nu_{p_i}(d), \beta_i\}=\max\{0, \nu_{p_i}(d)-\beta_i\}\qquad\text{for all $i\in \mathcal I$}.$$
Observe that $\theta_i\le \max\{0, \alpha_i-\beta_i\}$  for all $i\in \mathcal I$ and then $t$ divides $\dfrac {n_{\mathcal I}}{\gcd(n_{\mathcal I},q^R-1)}$.
Furthermore, if $\theta_i\ne 0$, then $\nu_{p_i}( d)= \theta_i+\beta_i\le \alpha_i$, and in the case $\theta_i=0$, we have $\nu_{p_i}(d)\le \alpha_i\le \beta_i$.
If follows that $d=d_0d_1$, where 
$$d_1=\prod_{i\in J} p_i^{\theta_i+\beta_j}=\gcd(k,n)\cdot\gcd(n_1, q^R-1),
\quad \text{with}\quad n_1= \prod_{i\in \mathcal J} p_i^{\alpha_i}$$
and $d_0$ is a divisor of $n_0=\prod_{i\in \mathcal I_0} p_i^{\alpha_i}$. Therefore, the number of $[n,k;q]$-codes is 
\begin{align*}
\frac 1k\sum_{d|n\atop \ord_d n=k} \varphi(d)&=\frac 1k\sum_{d_0| n_0} \varphi(d_0d_1)=\frac {n_0\cdot \varphi(d_1) } k\\
&=\frac {n_0\cdot\gcd(k,n) \cdot\gcd(n_1, q^R-1)} {k}\prod_{i\in \mathcal J} \left(1-\frac 1{p_i}\right).
\end{align*} 
By using the fact that $n_0=\gcd(n_0, q^R-1)$ and $\prod_{i\in \mathcal J} \left(1-\frac 1{p_i}\right)=\frac{\varphi(\gcd(k,n))}{\gcd(k,n)}$, we conclude that the number of irreducible cyclic $[n,k;q]$-codes is 
$$\frac {\gcd(n, q^R-1)\varphi(\gcd(k,n))} k.$$

On the other hand, by Lemma \ref{HOC},  the function $f(d)=\begin{cases} 1&\text{ if $d=1$}\\
\frac {\rho\cdot \varphi(d)}{\ord_d q}&\text{ if $d\ne 1$}\end{cases}$  is multiplicative for every $d$ divisor of $n$. So,
the total number of minimal cyclic codes of length $n$ is 
$$\sum_{d|n} \frac {\varphi(d)} {\ord_d q}= 1-\frac 1{\rho} +\frac 1\rho\sum_{d|n} f(d).$$
In order to calculate the sum, observe that
\begin{align*}
\sum_{d|p_i^{\alpha_i}} f(d)&=1+\sum_{s=1}^{\alpha_i} \frac {\rho\cdot(p_i^{s}-p_i^{s-1})} {\rho_i \frac {p_i^{s}}{\gcd(p_i^{s},q^{R}-1)}}\\
&=1+\frac{\rho}{\rho_i}\left(1-\frac 1{p_i}\right)\sum_{s=1}^{\alpha_i} \gcd(p_i^{s},q^{R}-1)\\
&=1+\frac{\rho}{\rho_i}\left(1-\frac 1{p_i}\right)\left[\sum_{s=1}^{\min\{\alpha_i,\beta_i\}} p_i^s +\max\{0, \alpha_i-\beta_i \}p_i^{\beta_i}\right]\\
&=1+\frac{\rho}{\rho_i}\left({p_i}^{\min\{\alpha_i,\beta_i\}}-1 +\left(1-\frac 1{p_i}\right)\max\{0, \alpha_i-\beta_i \}p_i^{\beta_i}\right)\\
&=1+\frac{\rho}{\rho_i}\left({p_i}^{\min\{\alpha_i,\beta_i\}}-1 +\max\{0, \alpha_i-\beta_i \}\varphi(p_i^{\beta_i})\right).
\end{align*}
Then, by using the fact that $\sum_{d|n} f(d)$ is a multiplicative function, we conclude the proof. 
\qed

\end{document}